\documentclass[5p,10pt]{elsarticle}

\usepackage{graphicx}
\usepackage{overpic}
\usepackage{amsmath} 
\usepackage{amssymb}
\usepackage{lineno}
\usepackage{url}
\usepackage{color}

\newcommand\beq{\begin{equation}}
\newcommand\eeq{\end{equation}}
\newcommand\beqa{\begin{eqnarray}}
\newcommand\eeqa{\end{eqnarray}}

\journal{ar$\chi$iv}

\begin{document}

\begin{frontmatter}

\title{Experiment and simulation of high-speed gas jet\\ penetration into a semicircular fluidized bed} 


\author[1]{William~D.~Fullmer}
\ead{william.fullmer@netl.doe.gov}
\cortext[cor1]{Corresponding author}
\author[2]{Jonathan~E.~Higham}
\author[1,3]{Roberto~Porcu}
\author[1]{Jordan~Musser}
\author[4,5]{C.~Wyatt~Q.~LaMarche}
\author[4]{Ray~Cocco}
\author[5]{Christine~M.~Hrenya}


\address[1]{National Energy Technology Laboratory, Morgantown, WV 26505, USA}
\address[2]{Department of Geography and Planning, School of Environmental Sciences, University of Liverpool, Liverpool L69 7ZQ, UK}
\address[3]{NETL Support Contractor, Morgantown, WV 26505, USA}
\address[4]{Particulate Solid Research, Inc., Chicago, IL 60632 USA}
\address[5]{Department of Chemical and Biological Engineering, University of Colorado, Boulder, CO 80309 USA}

%
%
\begin{abstract}
This work marks the third in a series of experiments that were in a semi-circular, gas-fluidized bed with side jets. In this work, the particles are (nominally) 1~mm ceramic beads. The bed is operated just at and slightly above and below the minimum fluidization velocity and additional fluidization is provided by two high-speed gas located on the sides of the bed near the flat, front face of the unit. Two primary measurements are taken: high-speed video recording of the front of the bed and bed pressure drop from a tap in the back of the bed. Particle Image Velocimetry (PIV) is used to determine particle motion, characterized as a mean Froude number, from the high-speed video. A CFD-DEM model of the bed is presented using the recently released MFIX-Exa code. Four model subvariants are considered using two methods of representing the jets and two drag models, both of which are calibrated to exactly match the experimentally measured minimum fluidization velocity. Although it is more difficult to determine the jet penetration depths in a straightforward manner as in the previous works using Froude number contours, the CFD-DEM results compare quite well to the PIV measurements, particularly for submodel {\texttt{flow\_Syam}}. Unfortunately, the good agreement of the solids-phase is overshadowed by significant disagreement in the gas-phase data. Specifically, the predicted time-averaged standard deviation of the pressure drop is found to be over an order of magnitude larger than measured. Due to the low value of the measurements, just 1\% of the mean bed pressure drop, it seems possible that the data is in error. On the other hand, the model may not be accurately capturing pressure attenuation through an under-fluidized region in the back of the bed. Without the possibility additional experiments to test the validity of the data, this work is simply being reported ``as is’’ without being able to indicate which, either the simulation or the experiment, is \emph{more} correct. 
\end{abstract}


\end{frontmatter}


\section{Introduction}
\label{sec.intro}
High-speed gas jets are frequently used in energy and chemical devices to provide additional fluidization in areas which are either under-fluidized or prone to jamming \cite{kunii, fan, merry71, hong97}. A set of experiments were carried out at PSRI studying jet penetration into a marginally fluidized bed of a semi-circular cross-section. Previous work focused on 6~mm plastic beads \cite{fullmer18a} and 3~mm ceramic beads \cite{fullmer20a}. This work extends the database to include a smaller diameter, nominally 1~mm, of the same ceramic material. Although still falling into the Geldart Group D classification \cite{geldart73}, the smaller 1~mm material is much closer to the Group B line, resulting in significantly more fluidization at similar jet velocities.

In addition to the experimental data, a focus in this work is on companion numerical modeling. The bed contains approximately 7.4 million particles. In order to simulate this experiment with a high-fidelity numerical model, we use the recently developed MFIX-Exa CFD-DEM code \cite{musser22, porcu23}. Built on AMReX \cite{zhang21, amrex}, MFIX-Exa is massively parallel, scalable, and performance portable.

In an effort to make this work reproducible, all of the measured data, simulation inputs and post-processing scripts are openly available. Small text files are stored in a version control github repository \cite{repo}. Large files, such as video and processed PIV matrices, are hosted on NETL's Energy eXchange Database (EDX), referenced within the repository.  

The remainder of this work is outlined as follows. In Sec.~\ref{sec.exp}, the experimental setup is reviewed, providing key geometric dimensions, flow conditions, and material characterization. The setup of the MFIX-Exa simulations are provided in  Sec.~\ref{sec.mfix} outlining the four models used to study the experiment. 
The results, namely bed pressure drop and particle velocimetry measurements, are presented in Sec.~\ref{sec.results} with a discussion on the comparison of simulation to experiment. The work is concluded in Sec.~\ref{sec.outro}.

\begin{table*}[htb]									
	\begin{center}								
	\caption{Summary of the measured dimensions of the experimental test section. (LL, LR, UL, and UR refer to the lower left, lower right, upper left and upper right jet locations.)}								
	\label{t.bed}								
	\begin{tabular}{lccc}								
		\hline							
		Bed property	&	Units	&	Measurement	&	Uncertainty ($\pm$)	 \\
		\hline							
		Width, $W$                       & (cm) & 28.58  &  0.1588  \\
		Maximum depth, $D_{mx}$	         & (cm)	& 15.169 &	$\sim$ 0.318  \\
		Jet diameter, $D_j$	             & (mm) &  3.86	 &	$<$ 0.01  \\
		right jet elevation, $y_{jL}$	 & (cm)	&  5.146 &	0.1588	\\
		left jet elevation, $y_{jR}$	 & (cm)	&  5.456 &	0.1588	\\
		left jet depth, $z_{jL}$ 	     & (cm)	&  1.667 &	0.1588	\\
		right jet depth, $z_{jR}$	     & (cm)	&  1.746 &	0.1588	\\
        pressure tap elevation, $y_{DP}$ & (cm) &  7.036 &  0.1588  \\
		\hline							
	 \end{tabular}								
	\end{center}								
\end{table*}									

\begin{figure}[htb]
\centering\includegraphics[width=0.96\linewidth]{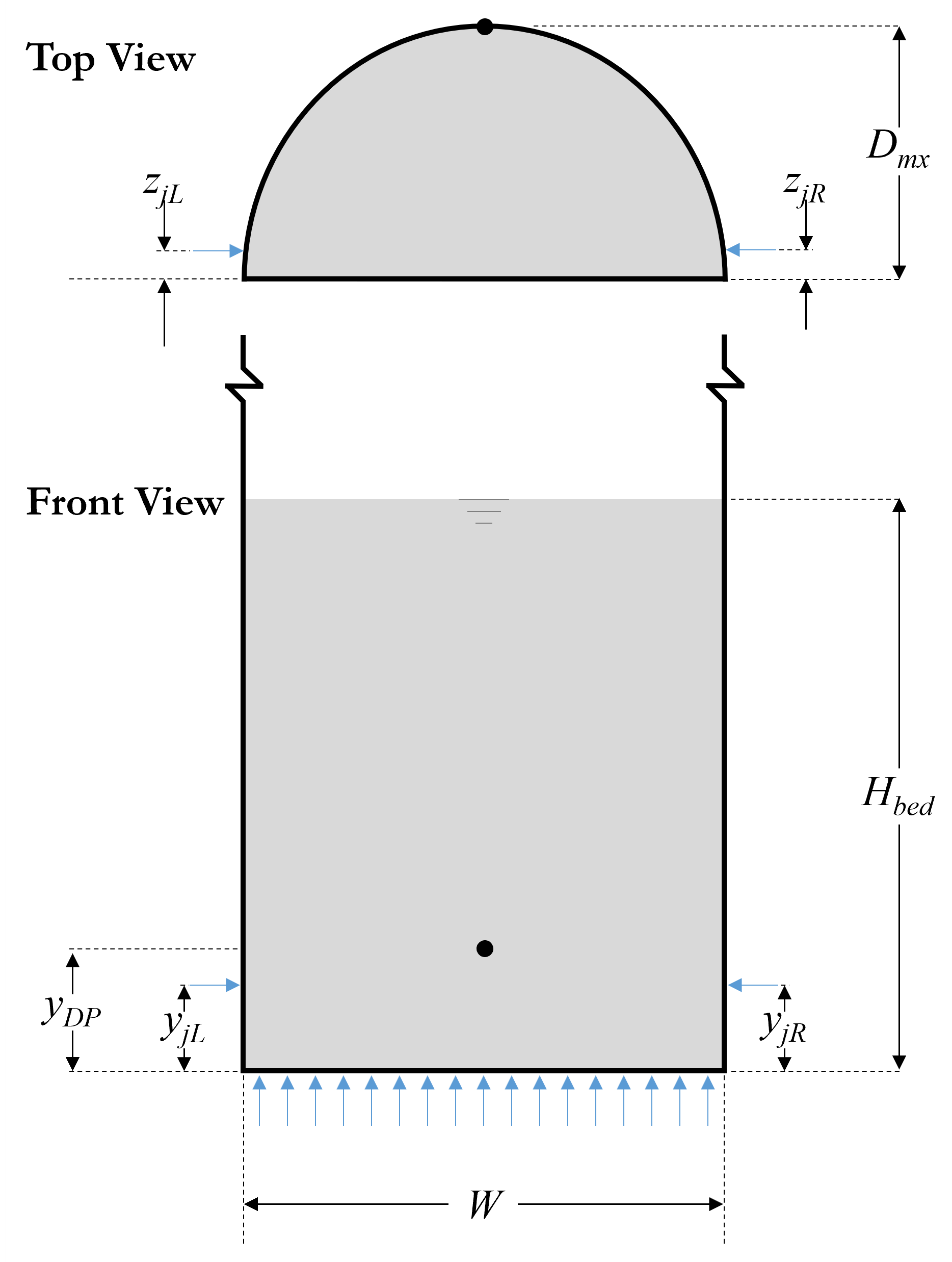}
\caption{(Color online.) Sketch of the test section showing key features with dimensions provided in Table~\ref{t.bed} (sketch not to scale).}
\label{fig.bed}
\end{figure}

\section{Experimental Setup}
\label{sec.exp}

\subsection{Bed description}
\label{sec.bed}
Sketched in Fig.~\ref{fig.bed}, the experimental test section is an acrylic fluidized bed approximately semi-elliptical in cross-section. The flat face is referred to as the front of the bed. A distributor fluidizes the bed from below near incipient fluidization, i.e., without the additional fluidization provided by the jets. The distributor is a drilled metal plate (0.089-inch diameter holes in a 5/32-inch triangular pitch) covered with a fine wire mesh. Two horizontal jets enter the bed near the face approximately $5$~cm above the distributor. (The experimental test section includes two additional jets \cite{fullmer18a}, but they are not considered in this work.) The dimensions of the bed are provided in Table~\ref{t.bed} with the measurement uncertainty. Interested readers should consult our previous publication \cite{fullmer18a} for a more detailed description of the complete unit and the test section that is the focus of this work.

\subsection{Material characterization}
\label{sec.particles}
The nominally 1~mm ceramic beads used in this work are characterized using the same methods as the nominally 3~mm ceramic beads \cite{fullmer20a} using methods originally developed by \citet{lamarche16, lamarche17}. Particle diameter, sphericity, density, restitution and friction coefficients are all measured. For reference, the mean values are listed in Table~\ref{t.particles} with uncertainty bands from a simple $t$-test. The raw data is provided in the project repository \cite{repo} along with script to extract key statics.

\begin{table*}[t]									
	\begin{center}								
	\caption{Summary of particle characterization measurements. (The raw data is available in the project repository \cite{repo}.)}								
	\label{t.particles}								
	\begin{tabular}{rcccc}								
		\hline							
		Property                       &  Symbol  &  Measurement &  95\% CI ($\pm$)  &  Units \\
		\hline							
		particle diameter              &  $d_p$       &  1.231   &  0.027  &  mm  \\
		sphericity                     &  $\psi$      &  0.9864  &  0.0011 &  --    \\
		restitution coeff.             &  $e_{pp}$    &  0.9165  &  0.0057 &  --    \\
		kinetic friction coeff.        &  $\mu_{pp}$  &  0.2173  &  0.0040 &  --    \\
		density                        &  $\rho_{p}$  &	\multicolumn{2}{c}{2611 - 2619}	&  kg/m$^3$  \\
		static bed height              &  $H_{bed}$   & 33.71    &  0.11   &  cm   \\
		Min. fluidization velocity     &  $U_{mf}$    & 53.25    &  2.54   &  cm/s \\
		\hline							
	 \end{tabular}								
	\end{center}								
\end{table*}		

After being loaded with material, the bed is fluidized without jets for several minutes to allow the particles to work into the (inactive) L-valve and equilibrate to a static height. Then, the flow is abruptly cut off and the static bed height, $H_{bed}$ is measured. This process is repeated for ten bed height measurements. The minimum fluidization velocity, $U_{mf}$ is then measured using a a set of Honeywell pressure transmitters (Model STD904-E1A-00000-AN.ZS.MB.1C+XXXX). Four pressure taps equipped with high-porosity brass snubbers (McMaster-Carr size 40-45 micron, Part No. 4034K2) are mounted along the back wall in the lower plenum, near the bottom of the bed, near the top of the bed and over a meter above the bed in the freeboard. The bed pressure drop, $DP_{bed} = p_g(y_{DP}) - p_{exit}$, is measured between an elevation of approximately $7$~cm above the distributor and the exit pressure, $p_{exit}$, over $1$~m above the top of the bed. The flow is increased in units of five standard cubic feet per minute (SCFM), the smallest demarcation on the flowmeter. This provides three points within the static bed region, two points are excluded because they are near minimum fluidization and three points in the well-fluidized region are recorded. The intersection of a quadratic fit to the under-fluidized points with the mean of the three well-fluidized points is taken as $U_{mf}$. The procedure is repeated three times giving $U_{mf} = 53.25 \pm 2.54$~cm/s.

\subsection{Operating conditions}
\label{sec.flow}
\begin{figure}[htb]
\centering\includegraphics[width=0.96\linewidth]{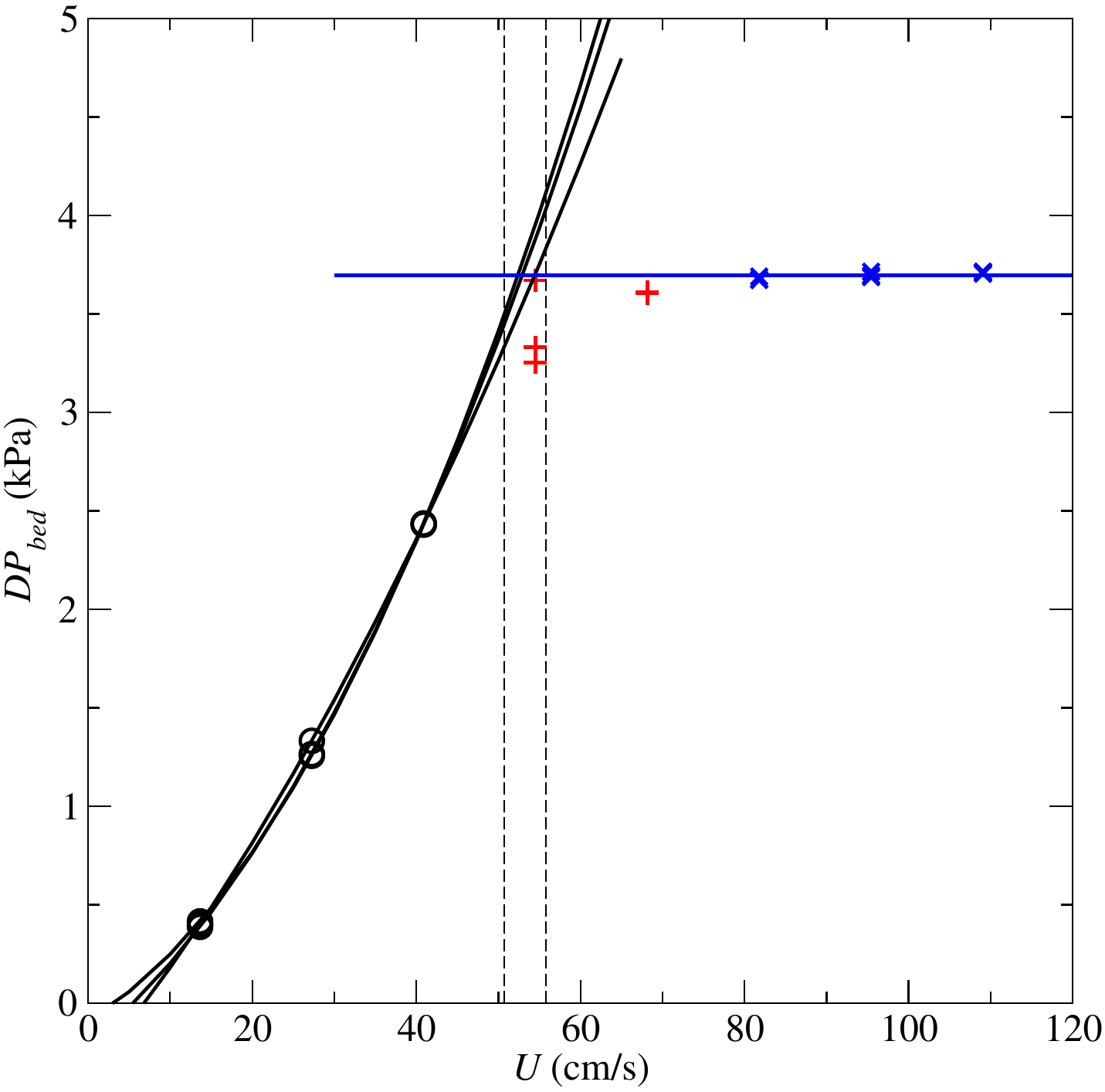}
\caption{(Color online.) Three traces of the minimum fluidization curve: the three lowest operating conditions (black $\circ$'s) are used to determine (individual) second-order polynomial curve-fits, the intermediate two operating conditions (red $+$'s) are neglected for being too near $Um_{mf}$, the three highest operating conditions (blue $\times$'s) are averaged to determine the fully fluidized bed pressure drop, the three intersections of the curves give $U_{mf} = 53.25 \pm 2.54$ (cm/s), the bounds of which are shown as vertical dashed lines.}
\label{fig.umf_exp}
\end{figure}

This work focuses on two bed operating conditions which are referred to as marginally fluidized and slightly fluidized. Five repeated experimental measurements are collected at each condition. The marginally fluidized case is operated right at minimum fluidization, $U/U_{mf} = 1.015$. The 1.5\% over-fluidization is well within the uncertainty on both the measured $U$ and calculated $U_{mf}$ values. Figure~\ref{fig.umf_exp} shows that the marginally fluidized operating condition would, without the aid of the jets, be within the hysteresis region and likely not fluidized well or completely. In the slightly fluidized operating condition, $U/U_{mf} = 1.27$ and the distributor flow appears to be just sufficient to fully fluidize the bed without the aid of the jets. At both conditions, the high-speed gas jets are operated at just under $200$~m/s. The measured velocities are summarized in Table~\ref{t.flow} with associated measurement uncertainties including uncertainties due to flow rate, area, back pressure and ambient temperature. 

\begin{table}[htb]													
	\begin{center}												
	\caption{Summary of bed operating conditions.}	
	\label{t.flow}												
	\begin{tabular}{rccc}												
		\hline											
		{ }	&	$U$ (cm/s)	&	$U_{j,\textrm{L}}$ (m/s)	&	$U_{j,\textrm{R}}$ (m/s) \\
		\hline											
		marginally	&	$54.05 \pm 8.71$	&	$197.0 \pm 11.7$	&	$197.0 \pm 11.7$ \\
		slightly	&	$67.63 \pm 8.83$	&	$197.2 \pm 11.8$	&	$197.2 \pm 11.8$ \\
		\hline											
	 \end{tabular}												
	\end{center}												
\end{table}

\section{Modelling}
\label{sec.mfix}
\subsection{Computational Model}
\label{sec.model}
In this work, we use the coupled computational fluid dynamics and discrete element method (CFD-DEM) to model the fluidized bed previously described. The CFD-DEM method has been outlined many times and will not be reiterated. Interested readers are referred to previous works \cite{deen07, garg12, garg12b, capecelatro13} for modeling details. For completeness, we note here that the CFD model uses a fluid grid that is of similar size to the particle but does not fully resolve fluid below the particle scale as in direct numerical simulation. A time-marching, soft-sphere method is used for the DEM model, i.e., particle collisions take place over a resolved number of timesteps.

\subsection{Numerical Method}
\label{sec.num}
The CFD-DEM model is solved with the finite volume numerical method embodied in the new MFIX-Exa code \cite{musser22, porcu23}. Built on the AMReX software framework \cite{amrex, zhang21}, MFIX-Exa was developed as part of the DOE's Exascale Computing Project (ECP) \cite{kothe18} suite of application codes to be deployed on the types of distributed, heterogeneous architectures found in leadership class computing facilities. The numerical method employees an operator splitting approach in which the particle and fluid work is largely segregated. The fluid phase is first advanced from time $t^n$ to $t^{n+1}$ using a projection method with a Godunov scheme. The Godunov method is used to extrapolate cell centered data at time $t^n$ to face centroids at $t^{n+1/2}$ using an intermediate marker and cell velocity field. An approximate projection method is applied to enforce the divergence constraint at the new time. The Lagrangian particles are then advanced using a standard forward Euler discretization with subcycling, i.e., using a separate $dt_p$ which is often smaller than the fluid step $dt_f = t^{n+1} - t^n$. A trilinear transfer kernel is used to deposit Lagrangian particle data onto the Eulerian fluid field and interpolate back to the particle. Irregular boundaries are modeled with embedded boundaries (EB). The intersection of the EB with the domain produces cut-cells in the fluid mesh and a level-set function for particle interactions. A state redistribution method is used to stabilize the fluid solve. Additionally, both the Godunov method of the fluid advance and the trilinear kernel of the coupling step must be appropriately augmented in the vicinity of an EB. The details of this numerical method have been reported in our previous works, specifically see \citet{musser22} for details of the over all algorithm and the EB-aware transfer kernel, see \citep{porcu23} for details of the EB-aware Godunov method, and see \citet{giuliani22} for details of the state redistribution method.

\subsection{Geometry}
\label{sec.prob}
Unlike the previously published datasets from this fluidized bed \cite{fullmer18a, fullmer20a}, the particle diameter of this material is smaller than the jet diameter so that the jets may be resolved by the fluid mesh. Taking that approach, the fluid mesh is set to $dx = 1.71$~mm so that $\sqrt{\pi/4}D_j = 2 dx$, i.e., the jet cross-sectional area is equivalent to four (uncut) fluid cells. OpenSCAD is used to construct and render the constructive solid geometry files used by MFIX-Exa to create the EB. The geometry of the fluidized bed is resolved with an ellipse clipped by a vertical plane. The vertical plane is aligned with the domain boundary at $z = 0^+$, shifted inside the domain a small tolerance. The jets are modeled in two different ways. First, the geometry of the jets is resolved. The walls of the jets are modeled as square ducts with an edge length of $3dx$. The jets are centered on the fluid mesh at nearest nodal location to the measured positions listed in Table~\ref{t.bed}. Consequently, in the computational model, $y_{jL} = 30 dx = 5.13$~cm, $y_{jR} = 32 dx = 5.47$~cm, and $z_{jL} = z_{jR} = 10 dx = 1.71$~cm. Note that while the jet elevations are within the measurement uncertainty and the L/R asymmetry is modeled (actually slightly accentuated), the L/R asymmetry in the jet depth is lost. However, this was necessary to ensure that the jets would enclose four uncut cells. The jets are extended to the domain extents where typical mass inflow boundary conditions (BC) are applied. In the second modeling approach, the jet walls are not resolved at all. Rather, the jet is modeled with a ``flow-on-EB'' type boundary condition applied to four cells on the EB modeling the bed wall. The boundary flow is imposed normal to the $x$-dimension even though the cut-cells on which the BC is applied are not. These two jet modelling approaches are referred to as {\tt EB} and {\tt flow} throughout. In both cases a standard mass inflow BC is applied at the inlet $y = 0$ and a pressure outflow at the exit, $y = L_y$. The domain size is $L_x = 2 L_z$, $L_y = 4 L_z$ and $L_z = 16.416$~cm resolved by a $192 \times 384 \times 96$ fluid mesh. Although MFIX-Exa is optimized for GPU accelerators, the majority of the compute resources on NETL's Joule 2.0 supercomputer are currently CPU-only. Therefore, the domain is partitioned into relatively small $16^3$ fluid grids, $2 \times 4 \times 6^3 = 1728$ in total, each grid managed by one MPI rank assigned to a single CPU. 336 grids contain only cells that fall outside of the active computational domain (i.e., cells covered by the boundary), therefore they are removed from the computation entirely. Of the remaining 1392 grids, 768 are cut by the EB and 624 are regular.

\subsection{Particles}
\label{sec.p}
The modeled particles are given the properties measured and reported in Table~\ref{t.particles}. Due to the fairly tight distribution, they are treated as monodisperse with the measured mean diameter. The mean density is also applied uniformly. The restitution and friction coefficients are applied to both particle and wall interactions. Sphericity is neglected as discussed below in Sec.~\ref{sec.drag}. It was assumed {\textit a priori} that the static bed packing (not measured) would be approximately 64\%, resulting in nearly 7.4 million particles. To measure the bed height, the bed was fluidized well above $U_{mf}$ for $2$~s and abruptly defluidized, similar to how $H_{bed}$ was measured in the lab. One single test resulted in approximately four to five layers of particles resting above the upper bound of the measured bed height. However, during fluidization tests (discussed below) the bed would further compact down to the lower bound of the measured $H_{bed}$. (Although this could have happened in the lab as well, it was not observed and the $U_{mf}$ was measured differently.) For this reason, we have opted to use the initial approximation of $N_p = 7.4 \times 10^6$ particles.

MFIX-Exa uses a simple linear-spring dashpot model for particle collisions. It is commonplace to use a small spring constant as long as the collision time remains much smaller than particle transport time scales, i.e., so that collisions appear to be nearly instantaneous. In this case, there is some concern that excessive overlap may happen in the near jet region due to impact of the high-speed gas jet into the dense particle bed. The spring constant is set to $k_n = 5 \times 10^5$~N/m, resulting in a particle-particle collision time (duration) of $\tau_{coll} = 5 \times 10^{-6}$~s. The particle subcycling time step is set to $dt_p = \min \left(dt_f,\; \tau_{coll}/20 \right)$. We close by noting that the low-$Ma$ projection method of MFIX-Exa is $cfl$-limited and, due to the high-speed of the gas jet BCs, so that the very small $dt_p$ value is somewhat mitigated by a relatively small $dt_f$ value as well.

\subsection{Drag Calibration}
\label{sec.drag} 
The final particle property listed in Table~\ref{t.particles} is the measured minimum fluidization velocity, $U_{mf}$. Because this work focuses on the impact of high-speed jets on a particle bed very near $U_{mf}$, it is imperative that the state of incipient fluidization is matched in the computational model. While there are several ways to match $U/U_{mf}$ with the experiment, we believe it is preferred to maintain the experimental inlet velocity, $U$, and adjust the drag model to match $U_{mf}$. To determine $U_{mf}$ of the model, the bed is fluidized at $U = 70$ for $1$~s to randomize the initial hexagonal particle configuration, abruptly defluidized ($U = 0$) for $1$~s to allow the particles to settle into a packed bed, and then fluidized at a rate of $dU/dt = 10$~cm/s$^2$ for $10$~s. The computationally measured $U_{mf}$ value is taken as the $U$ when gas-pressure at the inlet plane reaches 99\% of $N_p m \left|\bf{g}\right|/A_{bed}$. An MFIX-Exa native data reduction monitor is used to average the gas-phase pressure. To account for how the bed wall is resolved with the EB, the bed cross-sectional area, $A_{bed}$, is also computed with a monitor.

Two drag models have been considered: Wen and Yu \cite{wen66b} and Syamlal and O'Brien \cite{syamlal98}. Both models show a very linear response to their tuning parameters, therefore only four simulations are run (each) within the region of interest and calibration is determined with a least-squares fit. The Wen and Yu \cite{wen66b} drag model does not naively include a tuning parameter and, without any, results in a $U_{mf}$ almost 20\% below the measured value. Calibration is performed by including sphericity, i.e., modifying the particle diameter used in the drag model. The artificial sphericity needs to be increased to a value of $\Psi = 1.167$ to calibrate the modeled $U_{mf}$ with the experimental value. We note, again, that this is purely a calibration exercise as the measured sphericity was less than unity, as reported in Table~\ref{t.particles}.

Conversely, the Syamlal and O'Brien \cite{syamlal87} drag model uses a tuning parameter, $c_1$ (and $d_1 = 1.28 + \log c_1/ \log 0.85$). The parameter is traditionally set by providing the particle properties and the experimental $U_{mf}$, which results in a value of $c_1 = 0.788$. However, when included in the CFD-DEM model resulted in a significantly higher $U_{mf}$. For this model, the drag force has to be increased by decreasing $c_1$, finding a calibrated result of $c_1 = 0.4757$. Simulations run with the calibrated Wen and Yu \cite{wen66b} and Syamlal and O'Brien \cite{syamlal87} drag models are denoted {\tt WenYu} and {\tt Syam}, respectively. Each drag model is applied to each jet representations giving four model variants: {\tt EB\_WenYu}, {\tt EB\_Syam}, {\tt flow\_WenYu}, and {\tt flow\_Syam}. Each simulation is run for eight seconds of simulated time, the first two seconds is discarded as start-up transient.

\section{Results}
\label{sec.results}
\subsection{Bed pressure drop} 
\label{sec.DP}
The bed pressure drop, $DP_{bed}$, is measured from an elevation of $y_{DP} = 7.036 \pm 0.1588$~mm in the bed to the exit pressure. In the laboratory, the exit pressure tap is measured in the freeboard located $1.66$~m above the distributor. The bed pressure drop is recorded for a period of $T = 30$~s at a frequency of $F_s = 100$~Hz. The pressure measurements are ensured to encompass the shorter video measurements (see Sec.~\ref{sec.vp} below) but the exact relation between the two measurement windows is unknown, i.e., the pressure and video signals are not be time-synced. 

\begin{figure}[htb]
\centering\includegraphics[width=0.96\linewidth]{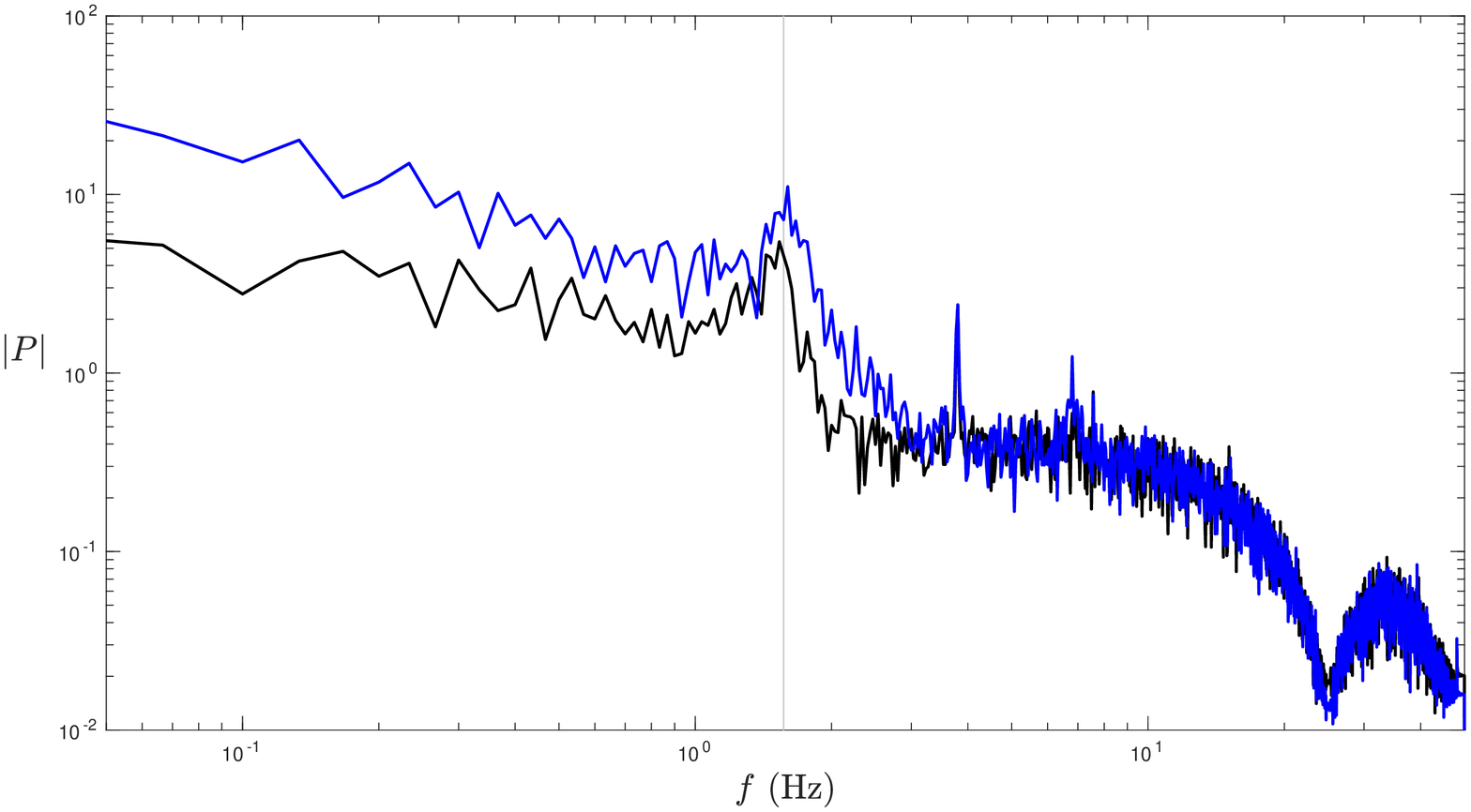}\\
\centering\includegraphics[width=0.96\linewidth]{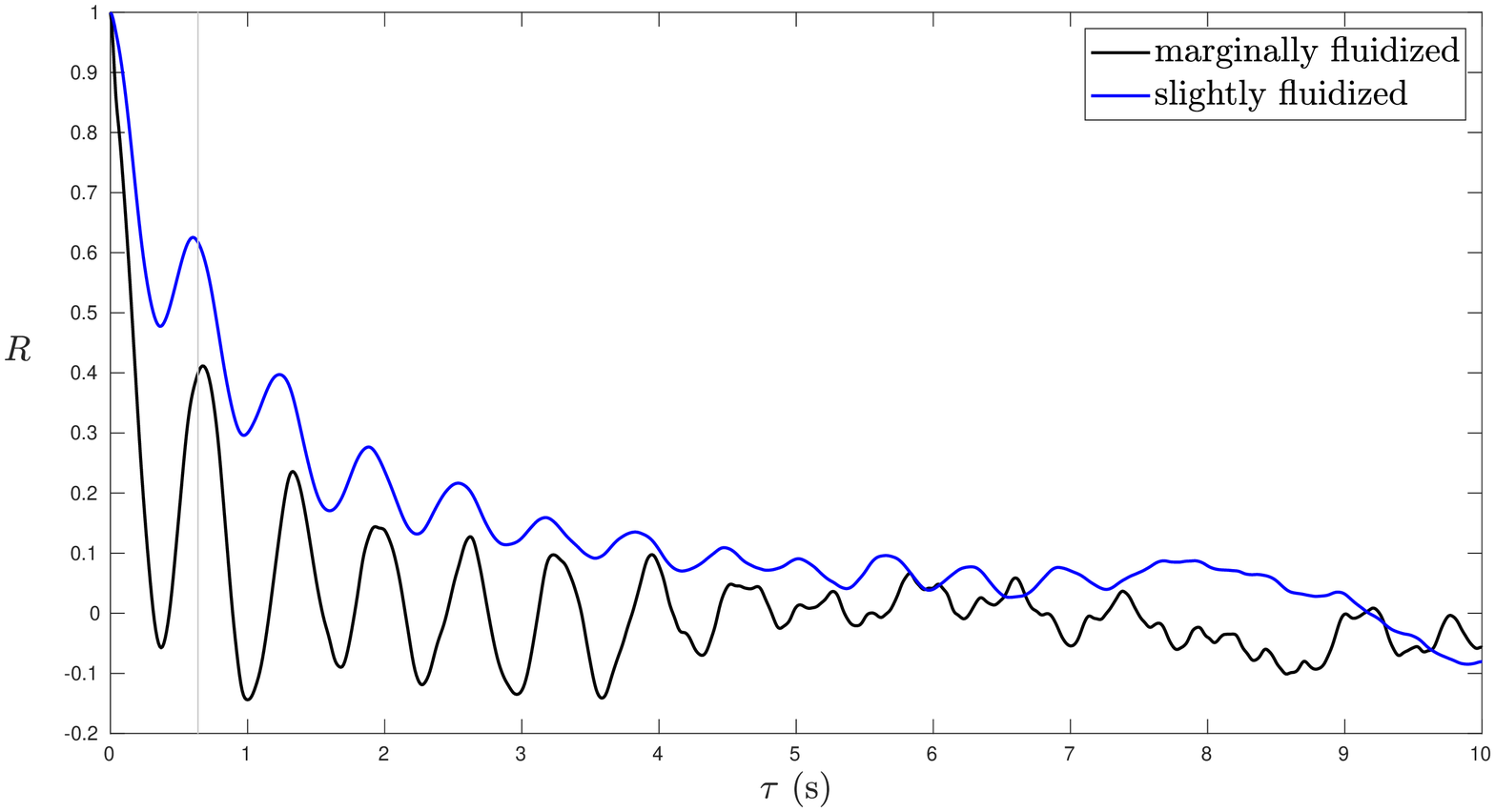}
\caption{(Color online.) Replicate averaged single-sided amplitude spectrum (top) and autocorrelation function (bottom) of the bed pressure drop at the marginally fluidized (black) and slightly fluidized (blue) operating conditions.}
\label{fig.fft_acf_exp}
\end{figure}

The frequency space of the bed pressure drop is explored first with the full $30$~s $DP$ signals. The single-sided power spectrum of the Fourier transform of the pressure drop is given by, 
\begin{equation*}
|P| = 2\left|\mathcal{F}_t\left(DP(t) - \overline{DP}\right)\right| / F_s T\; , 
\end{equation*}
where $\overline{DP}$ is the time-averaged value for each replicate measurement. MATLAB's {\tt fft} function is used to compute discrete Fourier transform. The replicate-averaged power spectrum, i.e., the average of the five spectra in frequency space, is shown in Fig.~\ref{fig.fft_acf_exp}. Originally, the signals exhibited a significant amount of high-frequency noise. Therefore, a very simple, sliding five-point trapezoidal filter is first applied to signals. The smoothing filter produces the distinct minima in the power spectra between $20$ and $30$ Hz. In both the marginally and slightly fluidized operating conditions, there is a noticeable local maxima in the spectra at $f_{bub} = 1.566$~Hz caused by the bubbling from the jets. The autcorrelation coefficient for integer $k$ from $0$ to $K < F_s*T$ is 
\begin{equation*}
R(k) = \frac{{\sum_{i=1}^{F_s*T-k} \left( DP(i) - \overline{DP}^{(-)}\right)\left( DP(i+k) - \overline{DP}^{(+)} \right)}}{{(F_s*T-k)DP'^{(-)}DP'^{(+)}}}\; , 
\end{equation*}
where $DP'$ is the time-averaged standard deviation and superscripts $(-)$ and $(+)$ indicate that time-averages have neglected $k$-values from the end or beginning, respectively, of the signal. The time lag is $\tau = k F_s$. The autocorrelation function is shown on the bottom of Fig.~\ref{fig.fft_acf_exp} for a time lag up to $\tau = 10$~s. The grey vertical line shows $\tau_{bub} = 1/f_{bub}$. The bubbling behavior is obvious for at least six periods. The bubbling frequency appears to dominate the signal of the marginally fluidized bed more than the slightly fluidized bed. In both cases, the signals become weakly correlated around six seconds, indicating that each replicate can be segmented into five samples. The twenty five samples (five replicates, five segments each) are used to compute an overall mean and standard deviation provided in Table~\ref{t.stats} with $t$-test confidence intervals (CIs) of 95\%. 

\begin{table*}[t]													
	\begin{center}												
	\caption{Comparison of time-averaged bed pressure drop and near-face particle velocity metrics between the (sample averaged) experiments and the MFIX-Exa CFD-DEM models.}
	\label{t.stats}												
	\begin{tabular}{lccccc}		
        \hline
        marginally fluidized & $\overline{DP}$ (Pa) &  $DP'$ (Pa)       &  $P_{jL}$ (cm)   &  $P_{jR}$ (cm)   & $\overline{Fr}$   \\
		\hline
        experiment           & $3563.3 \pm 3.6$     &  $21.83 \pm 0.66$ &  $8.92 \pm 0.13$ &  $8.95 \pm 0.08$ & $0.495 \pm 0.017$ \\
        {\tt EB\_WenYu}      & $3849.4$             & $293.4$           &  $7.13$          &  $7.01$          & $0.227$           \\
        {\tt EB\_Syam}       & $3906.7$             & $308.2$           &  $8.57$          &  $8.24$          & $0.660$           \\
        {\tt flow\_WenYu}    & $3856.0$             & $240.3$           & $13.06$          & $12.90$          & $0.238$           \\
        {\tt flow\_Syam}     & $3882.2$             & $271.1$           & $11.87$          & $11.73$          & $0.383$           \\
        \hline
        slightly fluidized   & $\overline{DP}$ (Pa) &  $DP'$ (Pa)       &  $P_{jL}$ (cm)   &  $P_{jR}$ (cm)   & $\overline{Fr}$   \\
        \hline
        experiment           & $3680.6 \pm 13.75$   &  $41.72 \pm 2.13$ & $11.89 \pm 0.50$ & $11.38 \pm 0.53$ & $1.151 \pm 0.113$ \\
        {\tt EB\_WenYu}      & $4208.7$             & $502.2$           &  $8.36$	       &  $8.08$          & $0.776$           \\
        {\tt EB\_Syam}       & $4288.6$             & $714.4$           & $10.82$          & $10.13$          & $1.395$           \\
        {\tt flow\_WenYu}    & $4088.1$             & $458.1$           &  -               & -                & $0.559$           \\
        {\tt flow\_Syam}     & $4148.4$             & $592.4$           &  -               & -                & $1.591$           \\
		\hline											
	 \end{tabular}												
	\end{center}												
\end{table*}

Originally, two types of native data reduction monitors were used to collect gas-phase pressure from the MFIX-Exa simulations. In the first method, the pressure is averaged over a 2-D plane at the elevation of the experimental pressure tap, $y_{DP} = 7.036$~cm. Note that this is actually extracted from the cell center in which $y_{DP}$ resides, i.e., the data corresponds to $y = 7.0965$~cm in elevation. In the second method, the pressure is averaged over a small volume--approximately 18 cells, half which are cut by the EB--mimicking the volume of the wall drilled out in the back of the bed for the pressure tap. This method also captures the uncertainty, $1.588$~mm, in the measurement of $y{DP}$. However, the pressure signals extracted with the two types of monitors are virtually indistinguishable and only the small volume, ``pressure-tap'' type monitor is used here. 

\begin{figure}[htb]
\centering\includegraphics[width=0.96\linewidth]{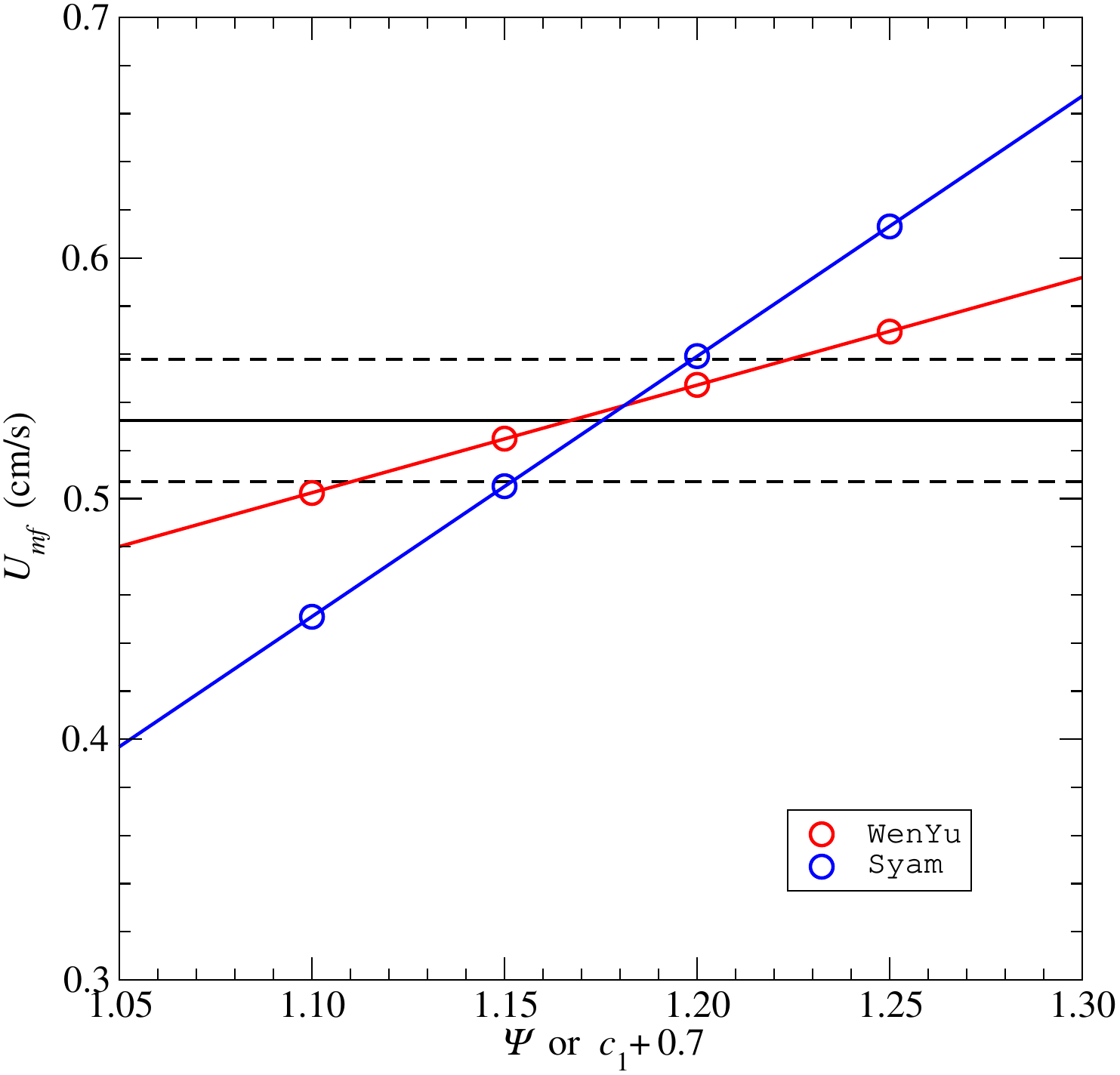}
\caption{(Color online.) Calibration of the Wen and Yu \cite{wen66b} (red) and Syamlal and O'Brien \cite{syamlal87} (blue) drag models to the experimentally measured minimum fluidization velocity $U_{mf} = 53.25$~cm/s (black line with dashes showing measurement uncertainty).}
\label{fig.umf_sim}
\end{figure}

\begin{figure}[htb]
\centering\includegraphics[width=0.96\linewidth]{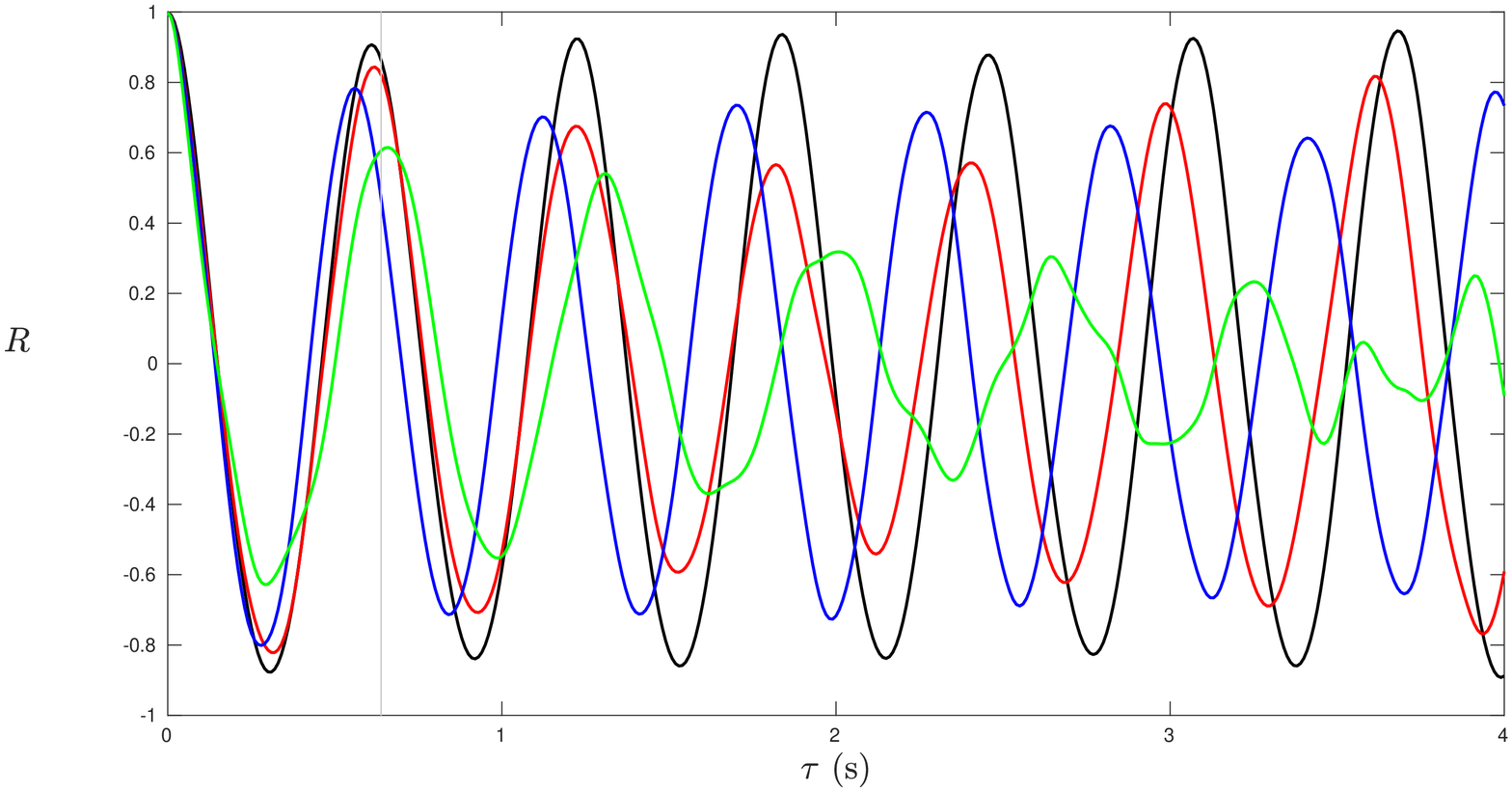}\\
\centering\includegraphics[width=0.96\linewidth]{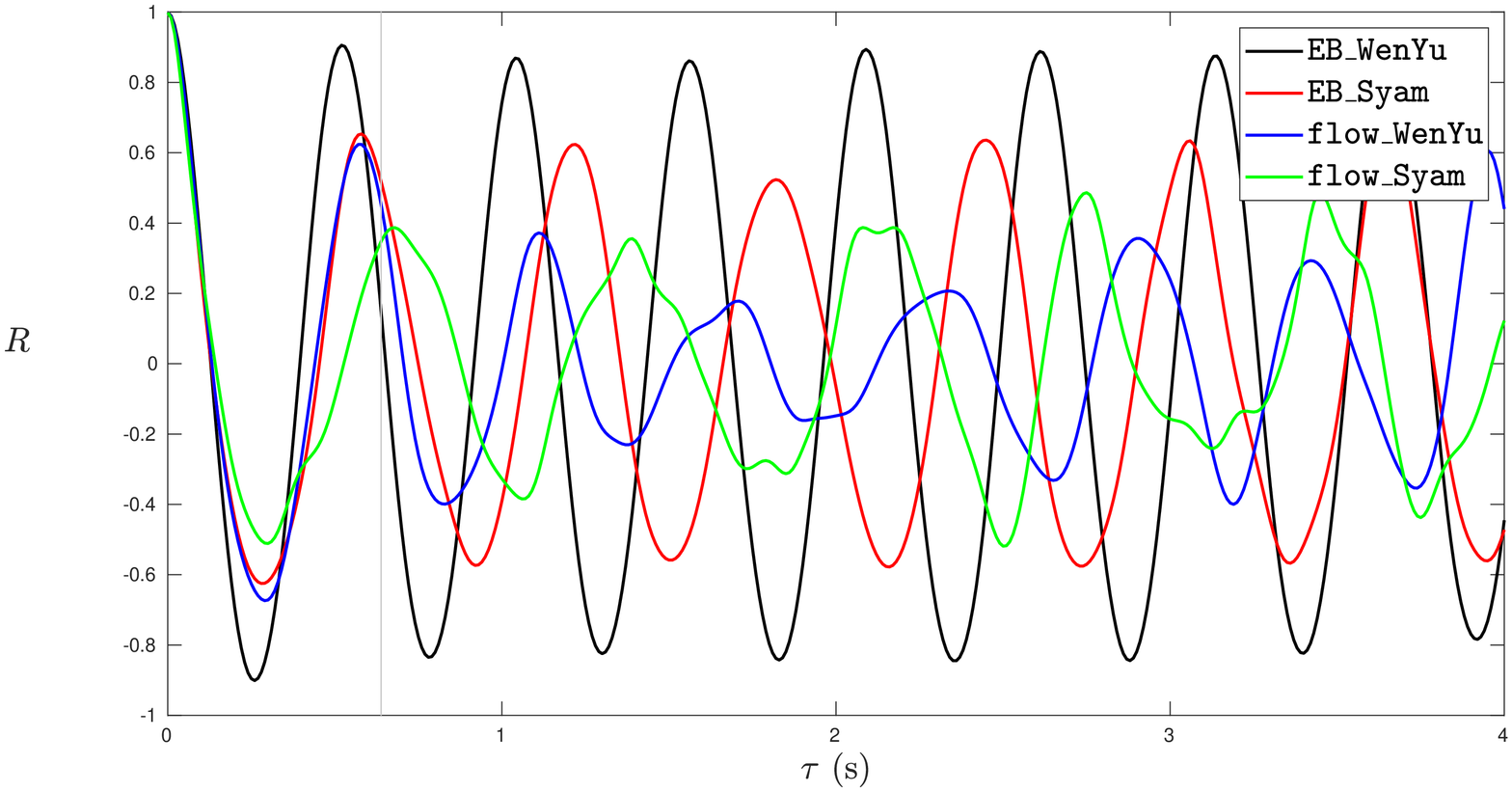}
\caption{(Color online.) Autocorrelation function of the {\tt EB\_WenYu} (black), {\tt EB\_Syam} (red), {\tt flow\_WenYu} (blue), and {\tt flow\_Syam} of the bed pressure drop at the marginally fluidized (top) and slightly fluidized (bottom) operating conditions.}
\label{fig.acf_sim}
\end{figure}

The autocorrelation function of the bed pressure drop signals for all four models at both operating conditions is shown in Fig.~\ref{fig.acf_sim}. The vertical grey line again corresponds to  $\tau_{bub} = 1.566^{-1}$~s. The correspondence between the peak frequency between the experiment and simulation is striking. Unfortunately, so is the regularity of the signal, which was not observed experimentally. For the magrinally fluidized system, three of the four models--{\tt flow\_Syam} being the exception--reach an anti-correlation minima greater than 80\% at a $\tau_{bub}/2$ shift and a correlation peak near or greater than 80\%  peak at a $1\tau_{bub}$ shift. This regularity is repeated, for the {\tt EB\_WenYu} model in particular, for at least $6\tau_{bub}$ periods. The slightly fluidized operating condition is nearly as periodic for the {\tt EB\_WenYu} model. However, the other models, specifically {\tt flow\_WenYu} are slightly less regular. This is not a new nor entirely unexpected result. Yet, it is somewhat surprising: the numerical model considers over 7 million discrete particles in an irregularly shaped bed with two very high speed jets producing bubbling that is as periodic as one might expect to find in a model with orders of magnitude fewer degrees of freedom (e.g., in a 2-D two-fluid model simulation). 

Discrepancies between the experiment and simulation are even more alarming for the time-averaged quantities. The mean bed pressure drop is reasonable, under 10\% error for the marginally fluidized condition and between 10\% and 15\% for the slightly fluidized condition. On the other hand, the (time-averaged) standard deviation is off by an order of magnitude in all cases, errors greater than 1000\%. Although there is no known issue with the experimental setup in either the transducers or data acquisition system, it seems possible, perhaps probable, that the $DP'_{bed}$ measurement is not accurate. To support this assumption we simply note that $DP'_{bed}/\overline{DP}$ of just 1\% for a bubbling fluidized bed is not consistent with previous literature \cite{vanommen11}. However, it should be noted that bubbling in this bed does not happen uniformly: bubbling is strongly correlated with the jets in the front of the bed while the pressure tap is in the back of the bed. If pressure tap were surrounded by a static region, the immobile particles could cause a significant attenuation of the gas-phase pressure. Yet, what little supplementary information is available (from a hand-held camera quickly showing the back of the experiment) seems to indicate that there is bubbling and bulk particle motion near the pressure tap, at least in the slightly fluidized condition.

\subsection{Particle velocity} 
\label{sec.vp}
Similar to the other two published works from this experimental test section \cite{fullmer18a, fullmer20a}, high speed video (HSV) imaging was taken of the flat face of the bed with front lighting to capture the particle dynamics near the front wall. The HSV was recorded with a Vision Research Phantom v7.2 camera with an acquisition rate of 1000 frames per second (fps), an exposure time of 151.75 $\mu$s, a resolution of 800$\times$600 pixels. The regions of interest is cropped manually with visual queues. The alignment of the region of interest is slightly different for the two operating conditions and the details are provided in the project repository \cite{repo}. However, each estimate the width of the bed to be 660 pixels. The recordings are memory-limited at 17696 frames at the resolution and frame rate. As detailed in the project repository \cite{repo}, two intermediate frames are withheld and each video is cropped into 5898-frame segments treated as independent samples of nearly 5.9~s HSV.

The width of the bed, $W$, is used to size the cropped HSV which gives an image resolution of approximately 2.3 px/mm. The mean particle diameter is resolved by less than three pixels, which is insufficient for particle tracking velocimetry (PTV) methods which typically rely on distinguishing individual particles. Therefore, we opt to use the more robust Particle Image Velocimetry (PIV) technique to measure particle dynamics in the HSVs. PIV \citep{adrian2011particle} determines displacements of regions or interrogation windows by finding the maximum correlation of each window, typically in the Fourier domain, from one image to the next. Gaussian distributions are fitted to correlation peaks allowing for `sub-pixel' accuracy \citep{hart2000piv}. In this study, we use an in-house PIV code \cite{higham2017using} based on the PIVlab algorithm developed by  \citet{thielicke2014pivlab}. The PODDEM algorithm \cite{higham16} was applied in a post-processing step to remove outliers.

The resulting PIV measurement is not a particle property but intricately tied to the size of the interrogation window. In our previous work \cite{fullmer20a}, for which both PTV and PIV were feasible, it was shown that the discrepancy between mesh-averaged-PTV and PIV is small if the window size is approximately the size of a particle. Here, however, the interrogation window is set to $16\; \times\; 16$ pixels with a 50\% overlap. The width of the region of interest is covered by 81 windows and four pixels along the right hand side are neglected: $(81+1)\times16/2 = 656$ pixels. With an edge length of approximately $5.6 d_p$ the window has a non-negligible effect on the velocimetry data and it is crucial that a similar post-processing step be used on the simulation data to capture this artifact.

In the MFIX-Exa simulations, particles are filtered by depth and those with $z_p < 10 d_p$ are saved at a frequency of $1000$~Hz as binary AMReX plot files for post-processing. One of the most well-known issues imaging dense, opaque particulate systems is the unknown and non-uniform imaging depth \cite{weber21}, i.e., how far into the bed is captured by the HSV. We choose, with no strong justification, a constant depth of $z_p \le 2 d_p$ when post-processing particle data. Post-processing is very similar to the PIV algorithm used to determine the velocimetry measurements. The region of interest is decomposed into $40 \times 40$ non-overlapping windows spanning the full bed width and the inlet to an elevation that is the nearest integer non-overlapping window from the top of the frame in the experimental HSV. For each data file, particles are sorted onto the non-overlapping mesh and then averaged onto $79 \times 79$ windows with 50\% overlap. The overlapping window mesh is used to compute time averaged statistics and compare with the experimental PIV measurements. 

\begin{figure*}[t]
\centering\includegraphics[width=0.96\linewidth]{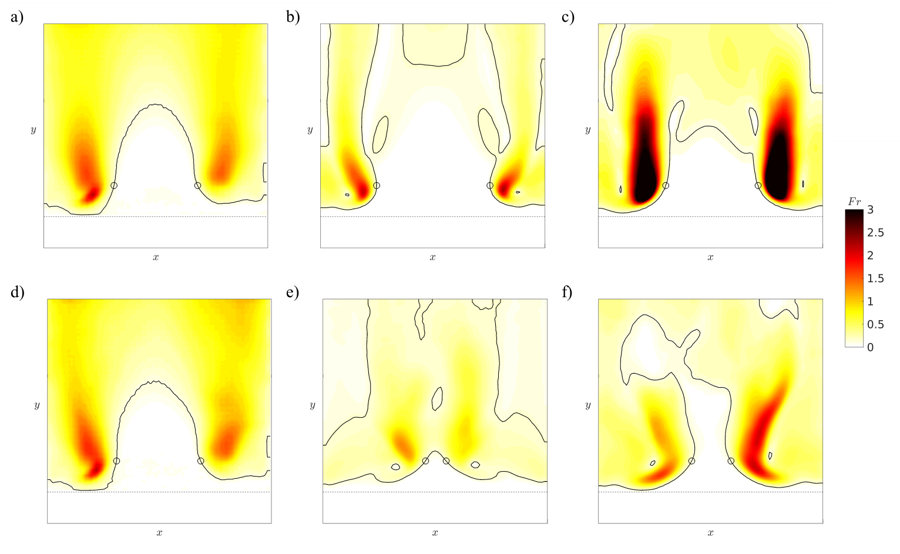}
\caption{(Color online.) Froude number contour plots of the marginally fluidized operating condition for a) the average of all PIV data, b) MFIX-Exa {\tt EB\_WenYu} model, c) MFIX-Exa {\tt EB\_Syam}, d) PIV of a single segment, e) MFIX-Exa {\tt flow\_WenYu} model, and f) MFIX-Exa {\tt flow\_Syam} model.}
\label{fig.Fr_marginally}
\end{figure*}

\begin{figure*}[t]
\centering\includegraphics[width=0.96\linewidth]{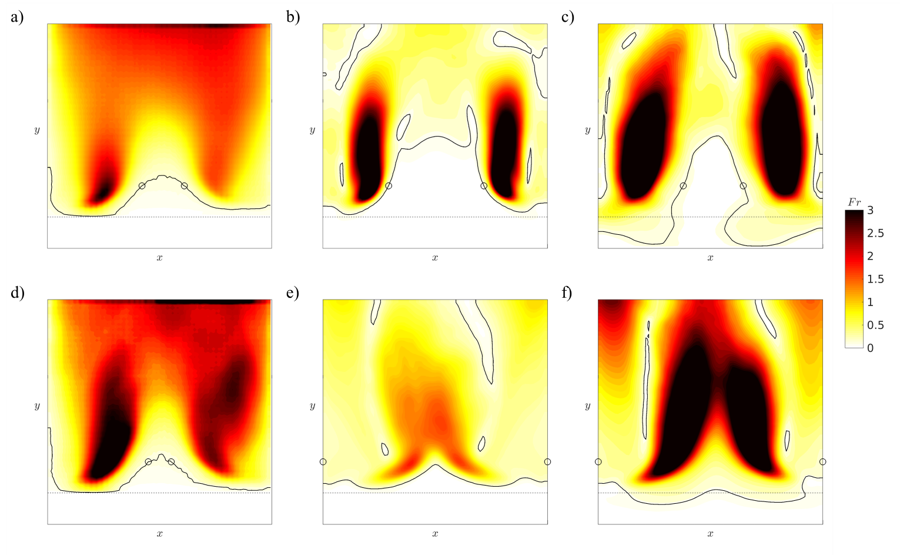}
\caption{(Color online.) Froude number contour plots of the slightly fluidized operating condition for a) the average of all PIV data, b) MFIX-Exa {\tt EB\_WenYu} model, c) MFIX-Exa {\tt EB\_Syam}, d) PIV of a single segment, e) MFIX-Exa {\tt flow\_WenYu} model, and f) MFIX-Exa {\tt flow\_Syam} model.}
\label{fig.Fr_slightly}
\end{figure*}

As in our previous works \cite{fullmer18a, fullmer20a}, the 2-D particle (spatially averaged solids) velocity field is non-dimensionalized into a Froude number, $Fr = \sqrt{(u^2 + v^2) /g d_p}$. Composite Froude number contour plots for the marginally and slightly fluidized operating conditions are provided in Figs.~\ref{fig.Fr_marginally} and \ref{fig.Fr_slightly}, respectively. One representative PIV figure is included and an average of all $15$ samples. All PIV results are available in the project repository \cite{repo}. The dashed black line at an elevation of $y = 40.259$~mm indicates the bottom of the frame of the experimental HSV. The region below this line down to the inlet is obscured by the flange which holds the distributor plate in place. 

Previously, the jet penetration depth, $P_j$, was measured as the furthest extant of the $Fr = 0.15$ isoline into the bed from the near-wall jet regions. The choice of $Fr = 0.15$ to distinguish between fluidized and static or under-fluidized regions still seems to be valid. Now, however, the jets are not isolated to two small, roughly circular, near-wall regions. The smaller material is more easily fluidized by the jets which produces a strong bubbling behavior. (Note that the jet velocity is approximately the same but the particle diameter is approximately one third compared to the other ceramic material studied \cite{fullmer20a}.) The ``furthest extant'' of the isoline criteria begins to breakdown because the isoline does not have a local maxima, rather increasing continuously to the centerline. There is still, at least for the marginally fluidized operating condition, a sizable static region in the center of the bed which makes the jet regions somewhat visually obvious. In an effort to facilitate comparison between experiment and simulation, we chose the simple, ad hoc solution of identifying $P_j$ as the lateral extent of the $Fr = 0.15$ isoline at an elevation of $y = 8$~cm, approximately $3$~cm above $y_j$. It should be noted that some generality in the measured jet penetration depth has been lost by this arbitrary definition. The left and right side jet penetration depths, $P_{jL}$ and $P_{jR}$, are identified by circles in Figs.~\ref{fig.Fr_marginally} and \ref{fig.Fr_slightly} and summarized in Table~\ref{t.stats}. The jet penetration depths of the sample-averaged $Fr$ plot are $P_{jL} = 8.94$~cm and $P_{jR} = 8.96$~cm for marginally fluidized and  $P_{jL} = 12.03$~cm and $P_{jR} = 11.14$~cm. All of which are very similar and within the CIs of the average of the penetration depths determined from single sample $Fr$ plots.

For both operating conditions and drag models, it is shown that the method of resolving the jets has a strong impact on the prediction of $P_j$. For the marginally fluidized operating condition, $P_j$ is under-predicted by the {\tt EB} jets and over-predicted by the {\tt flow} jets. The same is true of the slightly fluidized bed where the experiment shows a small unfluidized gap remaining and the {\tt flow} simulations predict a dynamic bed where the jets are fully merged at an elevation of $8$~cm. The choice of drag model also has an impact on the jet as well as the overall bed properties. The average Froude number, $\overline{Fr}$, can be used in addition to the jet penetration to measure the overall particle flow. (Note that for $\overline{Fr}$ calculated in from the simulations, data below $y = 40.259$~mm is withheld.) In all four cases, the {\tt WenYu} drag model under-predicts the experimentally measured (PIV) $\overline{Fr}$, while it is over-predicted in three of four cases by the {\tt Syam} drag model. This discrepancy is most likely due to the calibration, the drag force for {\tt WenYu} was turned down, the drag force for {\tt Syam} was turned up. While both calibrated drag models should provide a similar force for conditions near $U_{mf}$, such will not be the case throughout the bed, especially near the jets. It should be noted, though, that all models do a reasonable job at predicting the near-wall particle behavior. There was originally a concern that the $U_{mf}$-calibration method would produce entirely undesirable consequences and that does not appear to be the case, although there is still room for improvement.

\section{Summary and Outlook}
\label{sec.outro}
This work is the third, and likely final, in a series \cite{fullmer18b, fullmer20a} on opposing, high-speed gas jets penetrating horizontally into a bed of fluidized particles which would otherwise (i.e. without the jets) be near minimum fluidization. The bed is unique in that the cross-section is semi-circular (more accurately semi-elliptical) and the jets are located on the sides near the flat front face of the bed. This allows for clear imaging of the particle dynamics in the near jet region.

The bed is operated at a marginally fluidized condition, $U/U_{mf} = 1.015$, and a slightly fluidized condition, $U/U_{mf} = 1.27$. In both case, this refers only to uniform flow from the distributor below the bed. Additional fluidization is provided by two jets located approximately $5$~cm above the distributor and near the flat, front face of the bed operating at nearly $200$~m/s. Bubbles are created in the near-jet regions which travel inward as they rise through the bed. This leaves a small static or under-fluidized region in the center for the marginally fluidized condition which is all but eroded in the slightly fluidized condition. Experimental data is collected in the form of bed pressure signals at the back of the bed and high-speed video at the from of the bed. Particle Image Velocimetry (PIV) is used to measure solids (spatially averaged particle) velocity.

The bed is modeled with a relatively high-fidelity CFD-DEM model, resolving over 7 million particles and their collisions with a linear-spring-dashpot collision model using a stiff spring constant. The recently developed MFIX-Exa \cite{musser22, porcu23} is used for the simulations which consider four submodels: {\tt EB\_WenYu}, {\tt EB\_Syam}, {\tt flow\_WenYu}, and {\tt flow\_WenYu}, describing two types of jet representation and two drag models. Both drag models, Wen and Yu \cite{wen66b} and Syamlal and O'Brien \cite{syamlal87}, have been calibrated to match the experimentally measured $U_{mf}$ exactly.

Of the two types of data, the simulations do a much better job of capturing the arguably more difficult particle (solids) velocity. The jet and bulk bubbling behavior of all four models does a relatively good job of matching the PIV data, both qualitatively in terms of Froude number contour plots, see Figs.~\ref{fig.Fr_marginally} and \ref{fig.Fr_slightly}, and quantitatively in terms of jet penetration depths and mean velocity, see Table~\ref{t.stats}. The {\tt flow} models, which treat the jets as boundary conditions placed on the embedded boundary (EB) representing the bed wall, slightly over-predict jet penetration depths while the {\tt EB} models, which resolve the jets as internal flow with a containing EB of their own, slightly under-predict jet penetration. The difference may be attributed to the intersection of the two EB geometries, i.e., the bed and the jet, which will cause the jet cross-sectional area to slightly ``open up'' at the (jet) exit. There was a concern that adjusting drag to get a good results near minimum fluidization conditions could produce poor results at the near-jet and bubbling conditions, which, by and large, does not seem to be the case. However, the effect of the calibration can be seen in the results. The {\tt WenYu} models, for which the drag force was reduced, have a lower mean solids velocity, $\overline{Fr}$, than the {\tt Syam} models, for which the drag force was increased.

As encouraging as the particle velocity comparisons were, the gas-phase pressure drop comparisons were equally discouraging. The first problem is almost surely a modeling shortcoming: the pressure signals are almost perfectly periodic as shown by the autocorrelation function. It is worth noting that the predicted bubbling frequency does compare well with the experiment, the signal is just entirely too regular. This is perhaps not entirely unexpected, but the authors do find it somewhat surprising that such a complex and undeniably chaotic model can produce such periodic behavior. It is possible that relaxing some modeling simplifications, specifically uniform and constant particle properties: diameter, density, dissipation coefficients, etc., could produce a more realistic pressure signal.

Most concerning of all is the time-averaged standard deviation of the bed pressure drop. While this should be one of the most fundamental quantities, the comparison between simulation and experiment disagrees by over an order of magnitude. Because of the low $DP'_{bed}/\overline{DP}$ ratio of approximately 1\%, it seems possible that the data is incorrect, and, unfortunately, it is impossible to reevaluate at this point in time. It should be pointed out that the previously published work \cite{fullmer18a, fullmer20a} also report similarly low $DP'_{bed}/\overline{DP}$ values. However, it should be noted that the design of the experimental setup could be responsible for the low $DP'_{bed}/\overline{DP}$. Namely, the pressure tap is located in the very back of the bed while the jets produce bubbling at the front. Therefore, a counter hypothesis is that there is significant $\delta p_g$-attenuation in the experiment which is poorly captured by the CFD-DEM simulation. Unfortunately, at the time of this writing, it is nearly seven years since the data was collected and further investigation into this issue is not possible. Future work directed at the attenuation of gas-phase pressure signals from a well fluidized region through a static region could shed some light on this issue.

All of the raw data and particle property measurements are available in a project repository \cite{repo} that also contains post-processed data, both PIV and simulation, as well as the scripts used to post-process the simulation data. Access to the raw data and repository is available upon reasonable request from the corresponding author.

\section*{Acknowledgment}
The authors would like to thank Ann S. Almgren for developing the Godunov algorithm in MFIX-Exa, Deepak Rangarajan for developing the CSG-EB library, and Andrew Myers and Justin Weber for their help in developing the post-processing workflow. This computational research was supported by the Exascale Computing Project (17-SC-20-SC), a collaborative effort of the U.S. Department of Energy Office of Science and the National Nuclear Security Administration. The experimental work was supported by the U.S. Department of Energy under Grant No. DE-FE0026298. 


\bibliographystyle{model1-num-names}
\bibliography{wdf,higham,tmp}

\end{document}